\documentclass[aps,prb,twocolumn,showpacs,showkeys,floatfix]{revtex4}

\usepackage{graphicx}
\graphicspath{{figs/}}
\begin{document}
\title{Why edge effects are important on the intrinsic loss mechanisms of graphene nanoresonators?}
\author{Jin-Wu~Jiang}
    \altaffiliation{Present address: Institute of Structural Mechanics, Bauhaus of University, Marienstrae 15, 99423 Weimar, Germany. Electronic address: jwjiang5918@hotmail.com}
    \affiliation{Department of Physics and Centre for Computational Science and Engineering,
             National University of Singapore, Singapore 117542, Republic of Singapore }

\author{Jian-Sheng~Wang}
    \affiliation{Department of Physics and Centre for Computational Science and Engineering,
                 National University of Singapore, Singapore 117542, Republic of Singapore }

\date{\today}
\begin{abstract}
Molecular dynamics simulations are performed to investigate edge effects on the quality factor of graphene nanoresonators with different edge configurations and of various sizes. If the periodic boundary condition is applied, very high quality factors ($3\times10^{5}$) are obtained for all kinds of graphene nanoresonators. However, if the free boundary condition is applied, quality factors will be greatly reduced by two effects resulting from free edges: the imaginary edge vibration effect and the artificial effect. Imaginary edge vibrations will flip between a pair of doubly degenerate warping states during the mechanical oscillation of nanoresonators. The flipping process breaks the coherence of the mechanical oscillation of the nanoresonator, which is the dominant mechanism for extremely low quality factors. There is an artificial effect if the mechanical oscillation of the graphene nanoresonator is actuated according to an artificial vibration (non-natural vibration of the system), which slightly reduce the quality factor. The artificial effect can be eliminated by actuating the mechanical oscillation according to a natural vibration of the nanoresonator. Our simulations provide an explanation for the recent experiment, where the measured quality factor is low and varies between identical samples with free edges.
\end{abstract}

\maketitle

\pagebreak

\section{introduction}
Several recent works have investigated possible applications of the one-atom-thick graphene in the resonator field. The performance of resonators is described by the quality factor, which reflects the energy damping of the mechanical oscillation for the resonator. Both optical and electrical methods are applied to actuate the mechanical oscillation of graphene nanoresonators. Applying these two techniques, Bunch {\it et al.} detected quite low quality factors,\cite{BunchJS} while Zande {\it et al.} found in a recent experiment that the quality factor increases dramatically with cooling and can reach up to 9000 at 10 K with graphene resonators produced from the chemical vapor deposition growth method.\cite{Zande} A powerful technique in the nanoresonator field is to identify the shape of the mechanical oscillation of the nanoresonator by using a novel form of scanning probe microscopy.\cite{SanchezDG} This technique is useful for the detection of which natural vibration mode has been excited during the actuation of the mechanical oscillation of the nanoresonator. Particularly, it is useful to examine whether the mechanical oscillation of the nanoresonator is following a natural vibration of the system or not.

The quality factor is limited by both external and intrinsic damping mechanisms. Many efforts have been devoted to the investigation of various external damping mechanisms, such as charges in the substrate, attachment losses, and etc.\cite{SeoanezC} The intrinsic nonlinear effect was studied by a continuum elastic model derived from atomistic interaction for the graphene nanoresonator,\cite{AtalayaJ} or by molecular dynamics simulations.\cite{JiangJW} As another important intrinsic energy loss mechanism, the edge effect from free edges in the graphene was investigated by Kim and Park using molecular dynamics simulations in 2009\cite{KimSY}. They show that edge vibrations on a pair of free edges in the graphene accelerate the loss of the coherence of the mechanical oscillation of the nanoresonator, which results in very low quality factors. The recent experiment by Zande {\it et al.} also found that the edge effect is important for the variations in the resonance frequency and the quality factor of the graphene nanoresonator,\cite{Zande} which has not received adequate interpretation. This experiment has also found that free edges in the graphene can greatly reduce quality factors.

In recent years, much attention has been paid to the configuration of free edges in graphene nanoribbon. Gass {\it et al.} apply the high-angle annular dark-field technique to observe the edge reconstruction in graphene nanoribbon.\cite{GassMH} An unusual scrolling and staggering process was observed during the reconstruction. This unusual process is eventually attributed to the warping states, which are induced by intrinsic edge stress in graphene.\cite{ShenoyVB} Warping states are localized edge states with out-of-plane movement, and they are also found to be responsible for the edge reconstruction of the graphene nanoribbon under Joule heating treatment.\cite{JiaX,Engelund} Now, a natural question arises: is there any relation between these warping states and the reduction of quality factors in graphene nanoresonators by free edges?

In this paper, we report molecular dynamics simulations to study edge effects on graphene nanoresonators with both zigzag and armchair edges, and of various sizes. Very high quality factors are obtained for graphene nanoresonators with periodic boundary condition (PBC),
 \begin{figure}[htpb]
  \begin{center}
    \scalebox{1.0}[1.0]{\includegraphics[width=8cm]{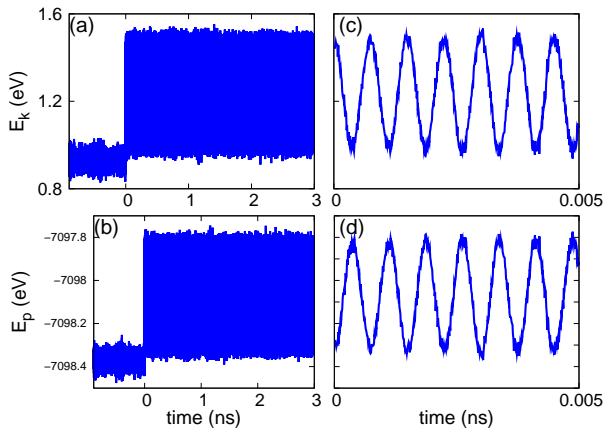}}
  \end{center}
  \caption{(Color online) Energy time history of Sample-A at 10~K with PBC for two short edges. (a) Kinetic energy. (b) Potential energy. (c) Close-up of kinetic energy. (d) Close-up of potential energy.}
  \label{fig_ek_pbc_ga830}
\end{figure}
 and the quality factor is inversely proportional to temperature as $1/T$, which is due to phonon-phonon scattering. However, quality factors have quite different value for zigzag and armchair graphene nanoresonators with free boundary condition (FBC). For zigzag edges, there are imaginary edge vibrations localized at free edges, which will flip between two doubly degenerate warping states during the mechanical oscillation of graphene nanoresonators. This flipping process breaks the coherence of the mechanical oscillation of the graphene nanoresonator and leads to extremely low quality factors with temperature dependence $1/T^{0.28}$. The quality factor of zigzag graphene nanoresonator will be further reduced with temperature dependence $1/T^{0.16}$, if the oscillation of the resonator is actuated according to an artificial vibration (non-natural vibration of the system). For armchair edges, there is no imaginary edge vibration, so quality factors are mainly reduced by the artificial effect. An extremely high quality factor can be obtained for armchair grphene nanoresonators if the mechanical oscillation is actuated naturally (according to its natural vibration), although FBC is applied.

The rest of the paper is organized as follows. Sec.II(A) is devoted to some techniques for the simulation and calculation approach. In Sec.II(B), the imaginary edge vibration effect on quality factors is discussed. The artificial effect on quality factors is illustrated in Sec.II(C). The paper is concluded in Sec.III.

\section{results and discussion}
\subsection{simulation setup}
\begin{figure}[htpb]
  \begin{center}
    \scalebox{0.8}[0.8]{\includegraphics[width=8cm]{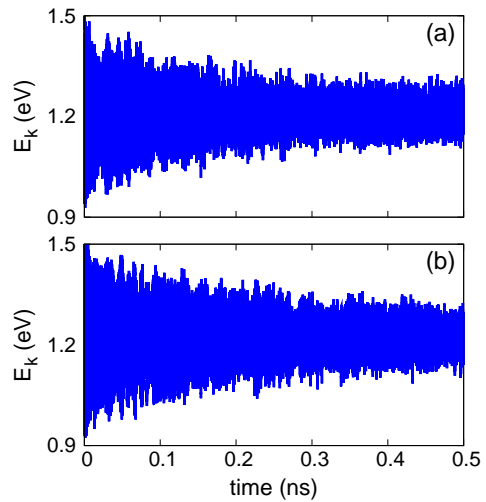}}
  \end{center}
  \caption{(Color online) Kinetic energy time history of Sample-A at 10~K with FBC. The mechanical oscillation is actuated artificially in (a) and naturally in (b).}
  \label{fig_ek_fbc_ga830}
\end{figure}
The first simulated graphene sample is of size $(L_{x}, L_{y}) = (19.7, 127.8)$~{\AA}, which will be referred to as Sample-A throughout the text. Short edges are two zigzag edges, so this type of graphene is usually called zigzag graphene nanoribbon. The x-axis is along the short edge, and the y-axis is along the long edge. The z-axis is perpendicular to the graphene plane. Two long edges are always fixed in all simulations. We apply either PBC or FBC for the two short edges. The interatomic interaction is described by the Brenner potential.\cite{Brenner} The Newton equations of motion are integrated using the velocity Verlet algorithm with a time step of 1 fs.

Fig.~\ref{fig_ek_pbc_ga830} shows the time history of kinetic energy in panel (a) and potential energy in panel (b) for Sample-A at 10 K. PBC is applied at the two short edges. For $t<0$, the N\'ose-Hoover\cite{Nose,Hoover} heat bath is applied to thermalize the system to a constant temperature within the NVT ensemble. In panel (a), there is a 5\% variation in the kinetic energy due to the statistical natural of the N\'ose-Hoover algorithm. At $t=0$, the nanoresonator is actuated by adding a sinuous velocity $\Delta\vec{v}_{j}\propto \vec{e}_{z} \sin (\pi x_{j}/L_{x})$ to each atom $j$, where $x_{j}$ is the $x$-coordinate of the atom and $\vec{e}_{z}$ is the unit vector in $z$ direction. The energy increase by the added velocity $\Delta\vec{v}$ is $50\%$ of the total kinetic energy. It is large enough to ensure that the mechanical oscillation of the nanoresonator is obviously stronger than the thermal vibration in the nanoresonator. Yet, it is less than $0.003\%$ of the total potential energy. For $t>0$, the nanoresonator oscillates within the NVE ensemble. Panel (c)/(d) is the close-up of the kinetic/potential energy. Panels (c) and (d) demonstrate the energy exchange between kinetic and potential energy after $t>0$, while the total energy is kept as a constant. The energy exchange between kinetic and potential energy reflects the mechanical oscillation of the nanoresonator. Some roughness can be observed on top of the curve in panels (c) and (d), which display the thermal vibration at 10~{K}. The decay of the oscillation amplitude is used to analysis the quality factor of the nanoresonator by fitting the kinetic energy after $t>0$ to a function $E_{k}(t)=a+b(1-2\pi/Q)^{t} \cos(\omega t)$. $\omega$ is the natural frequency of the nanoresonator.
\begin{figure}[htpb]
  \begin{center}
    \scalebox{0.9}[0.9]{\includegraphics[width=8cm]{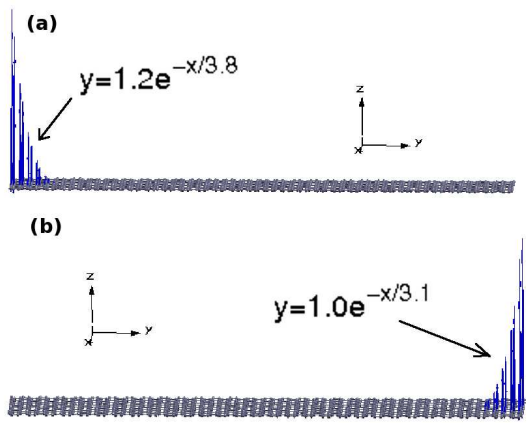}}
  \end{center}
  \caption{(Color online) Two edge vibrations localized nearby the two free short edges in Sample-A. The arrow attached on each atom respresents the vibrational amplitude of this atom in the vibration. Vibrational amplitudes decay exponentially from edge into center. The frequencies of these two edge vibrations in (a) and (b) are degenerate and imaginary, $\omega^{2}=-625$~{cm$^{-2}$}.}
  \label{fig_u_edgemodemode_ga830}
\end{figure}
 Parameter $a$ gives final kinetic energy after the mechanical oscillation dies away. The decay of the mechanical oscillation will lead to the temperature increase in the nanoresonator. $Q$ is the resulted quality factor. The symbol $Q$ will be used to denote the normal mode coordinate throughout the text. This is the only place where $Q$ denotes the quality factor. The obtained quality factor is as high as $3\times10^{5}$ at 10~{K}. Similar fitting procedure can be conducted for the potential energy, which results in a same value for the quality factor. In the following, the quality factor will be calculated from the fitting of the kinetic energy.

\subsection{effects from imaginary edge vibrations}
It was shown that the quality factor of a graphene nanoresonator will be dramatically reduced by free edges.\cite{KimSY} We repeat this result in Fig.~\ref{fig_ek_fbc_ga830}~(a) for Sample-A, which is also the sample studied by Kim and Park. FBC has been applied for the simulation, while all other parameters are the same as Fig.~\ref{fig_ek_pbc_ga830}. In particular, we would like to stress that the mechanical oscillation is also actuated by adding the sinuous velocity $\Delta\vec{v}$, which is the same as Ref.~\onlinecite{KimSY}. Our simulation demonstrates that the mechanical oscillation decays very fast in presence of free edges, resulting in an extremely low quality factor.

\begin{figure}[htpb]
  \begin{center}
    \scalebox{1.0}[1.0]{\includegraphics[width=8cm]{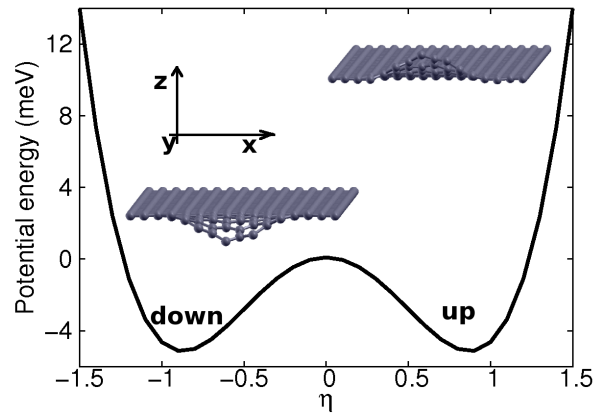}}
  \end{center}
  \caption{Brenner potential energy for graphene nanoribbon with position $\vec{r}'_{j}=\vec{r}^{0}_{j}+\eta \vec{\chi}_{\rm edge vibration}^{j}$ for atom $j$, where $\vec{r}^{0}$ is the optimized structure. $\vec{\chi}_{\rm edge vibration}$ is the eigen vector of the imaginary edge vibration localized at one short edge as shown in Fig.~\ref{fig_u_edgemodemode_ga830}~(a). The right top inset displays a configuration for $\eta >0$, where edge atoms are above the graphene plane. The left bottom inset demonstrates a configuration for $\eta <0$, where edge atoms are below the graphene plane.}
  \label{fig_imaginary_mode}
\end{figure}
The origin for this edge effect is still unclear, although the phenomenon of low quality factor due to free edges has been shown in previous work. To reveal the underlying mechanism, we study all natural vibrations in Sample-A with FBC by solving the eigenvalue problem of dynamical matrix derived from the Brenner potential. The dynamical matrix is obtained from $K_{ij}=\partial^{2}V/\partial x_{i}\partial x_{j}$, where $V$ is the Brenner potential and $x_{i}$ is the position of the i-th degree of freedom. This formula is realized numerically by calculating the energy change after displace the i-th and j-th degrees of freedom for a small value. We find several edge vibrations localizing at the two free edges. Fig.~\ref{fig_u_edgemodemode_ga830} displays two lowest-energy edge vibrations. The arrow attached on each atom represents the vibrational amplitude of this atom in the vibration. Vibrational amplitudes decay exponentially from edge into center, exhibiting the localization property. These two edge vibrations are degenerate. They localize at each end of the two free edges. An interesting property of these two edge vibrations is their imaginary frequency, $\omega^{2}=-625$~{cm$^{-2}$}. We have optimized the structure before the derivation of the dynamical matrix from the Brenner potential. During the structure optimization, the residue force on each atom has been required to be as small as $10^{-4}$ eV/\AA, so these two imaginary vibrations do not result from numerical errors; instead, they reflect the intrinsic properties of free zigzag edges.

\begin{figure}[htpb]
  \begin{center}
    \scalebox{1.0}[1.0]{\includegraphics[width=8cm]{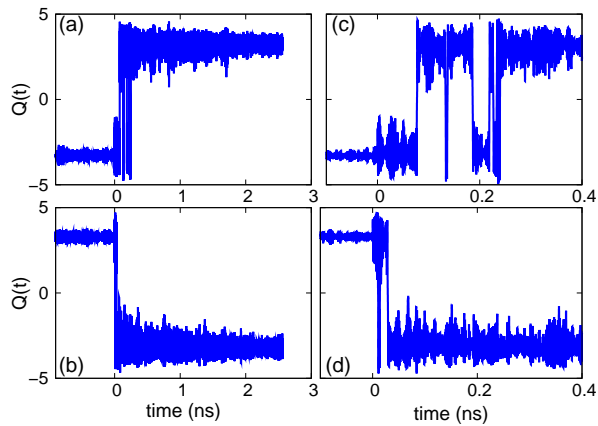}}
  \end{center}
  \caption{(Color online) Normal mode coordinate time history of the two edge vibrations in Fig.~\ref{fig_u_edgemodemode_ga830} showing flipping phenomena. The simulated system is Sample-A at 10~{K} with FBC. The mechanical oscillation is actuated artificially. (a) is for the edge vibration shown in Fig.~\ref{fig_u_edgemodemode_ga830}~(a). (b) is for the edge vibration shown in Fig.~\ref{fig_u_edgemodemode_ga830}~(b). (c) is the close-up for (a). (d) is the close-up for (b).}
  \label{fig_normalmode_artificial_ga830}
\end{figure}
Imaginary edge vibrations imply the instability of free zigzag edges when the graphene nanoribbon vibrates according to these two imaginary edge vibrations. Similar instability was also found by Lan {\it et al.} using tight-binding potential.\cite{LanJ} Starting with these imaginary vibrations, we can search for the ground state corresponding to a stable structure. Fig.~\ref{fig_imaginary_mode} shows the potential energy of the graphene nanoribbon with position $\vec{r}'_{j}=\vec{r}^{0}_{j}+\eta \vec{\chi}_{\rm edge vibration}^{j}$ for atom $j$, where $\vec{r}^{0}$ is the optimized structure. $\vec{\chi}_{\rm edge vibration}$ is the eigen vector of the imaginary edge vibration localized at one short edge as shown in Fig.~\ref{fig_u_edgemodemode_ga830}~(a). Similar potential curve is also found for the other imaginary edge vibration shown in Fig.~\ref{fig_u_edgemodemode_ga830}~(b). The parameter $\eta$ is in [-1.5, 1.5] in the figure. Two insets display the edge configuration for the graphene nanoribbon with $\eta>0$ (right top inset) and $\eta<0$ (left bottom inset). The potential curve demonstrates that the optimized structure (at $\eta=0$) actually corresponds to a local maximum value of potential energy instead of a minimum value. Meanwhile, there are two ground states at $\eta=\pm 0.9$, which are doubly degenerate to each other. We refer to these two ground states as $\eta_{\uparrow}$ and $\eta_{\downarrow}$ states, respectively. They are the warping states, where the graphene nanoribbon is scrolled and warped at the edge.\cite{ShenoyVB} The graphene nanoribbon has a mirror reflection symmetry, $\sigma_{xy}$, with respect to the $xy$ plane, resulting from its two-dimensional nature. Each of the warping state alone breaks this mirror reflection symmetry as can be seen from both insets. This is a spontaneous symmetry breaking phenomenon resulting from the Mexican-top-hat-like potential shown in Fig.~\ref{fig_imaginary_mode};\cite{DaviesPCW} i.e one symmetrical system ends up with an apparently asymmetric ground state, which is ubiquitous in condense matter physics such as superconductors\cite{NagaosaN} and superfluids,\cite{MorandiG}, and is related to the Nambu-Goldstone boson in the scalar field theory\cite{GoldstoneJ}. The mirror reflection symmetry can be restored if these two warping states are considered together, as the mirror operation transforms one warping state to the other. The optimization at 0 K preserves the mirror reflection symmetry in the graphene nanoribbon; thus it can not achieve either of the two ground (warping) states. The frequency of the imaginary vibration ($25i$~{cm$^{-1}$}) gives us some information about the energy barrier between these two warping states: $\hbar \omega=25/8.06554\approx 3.1$~{meV}. This value is comparable with the energy barrier, 5.0~{meV}, as reading directly from the potential curve.

\begin{figure}[htpb]
  \begin{center}
    \scalebox{1.0}[1.0]{\includegraphics[width=8cm]{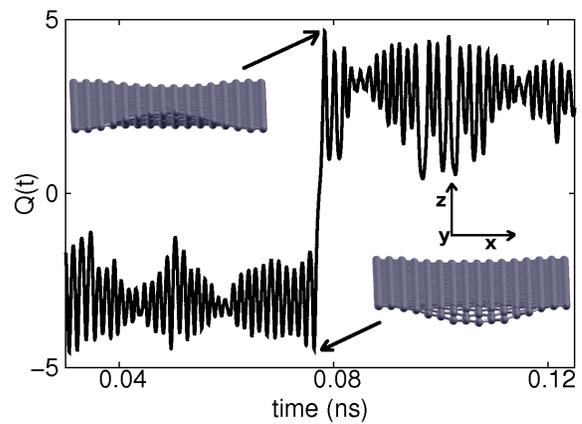}}
  \end{center}
  \caption{The edge configuration of the graphene nanoresonator undergoes a sudden change from $\eta_{\downarrow}$ to $\eta_{\uparrow}$ state during the first flipping process in Fig.~\ref{fig_normalmode_artificial_ga830}~(c).}
  \label{fig_flip}
\end{figure}
As the edge vibration manifests the characteristic difference between PBC and FBC, it is natural to attribute the extremely low quality factor to edge vibrations. To verify this thinking, we monitor the evolution of these two imaginary edge vibrations during the molecular dynamics simulation. The normal mode coordinate for a natural vibration $k$ can be obtained from its standard definition
\begin{equation}
Q_{k}(t)=\sum_{j=1}^{3N}\chi^{k}_{j}(r_{j}(t)-r_{j}^{0}),
\label{eq_normalmode}
\end{equation}
where $3N$ is the total degrees of freedom. $\vec{\chi}^{k}$ is the eigen vector for this natural vibration. $r_{j}(t)$ is the position of the $j$-th degree of freedom and $r_{j}^{0}$ is the equilibrium position. Time histories of $Q(t)$ for these two imaginary edge vibrations are shown in Fig.~\ref{fig_normalmode_artificial_ga830}. For $t<0$, the thermal process is controlled by heat bath and $Q(t)$ oscillates around a nonzero value, which is a more stable state. After the removal of heat bath for $t>0$, there is a flipping phenomenon; i.e the normal mode suddenly flips between two stable states. The magnitudes of $Q(t)$ are the same in these two stable states, while the signs are opposite. It indicates that energy does not change after the flipping process, since energy is proportional to $Q^{2}(t)$. We now justify that these two stable states are exactly the two warping states discussed above. The two insets in Fig.~\ref{fig_flip} show the edge configurations of the graphene nanoresonator before and after the flipping process. Indeed, the system jumps from the warping state $\eta_{\downarrow}$ to the other warping state $\eta_{\uparrow}$ during this flipping process. These two warping states are degenerate, so the energy does not change during the flipping process. This corresponds to the above fact that the $Q(t)$ changes sign with unchanged amplitude during the flipping process. The flipping process only exists in the early time history, when the system is out of thermal equilibrium. There is no flipping process after the mechanical oscillation dies away; i.e the whole system reaches thermal equilibrium again.

\begin{figure}[htpb]
  \begin{center}
    \scalebox{0.75}[0.75]{\includegraphics[width=8cm]{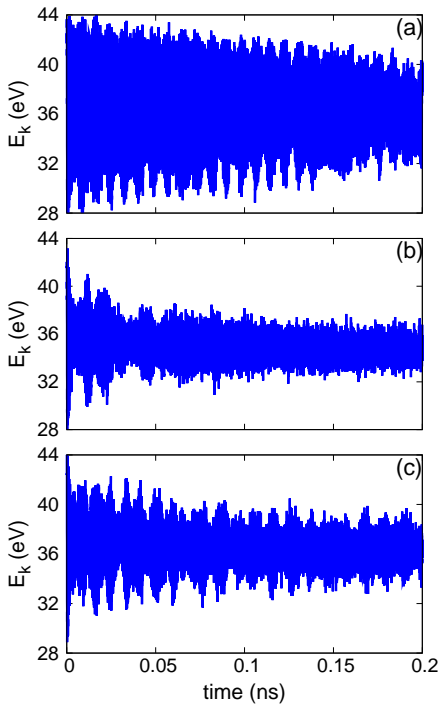}}
  \end{center}
  \caption{(Color online) Kinetic energy time history at 300~{K} for Sample-A with (a) PBC, (b) FBC and artificial actuation, and (c) FBC and natural actuation.}
  \label{fig_ek_300K_ga830}
\end{figure}
\begin{figure}[htpb]
  \begin{center}
    \scalebox{1.0}[1.0]{\includegraphics[width=8cm]{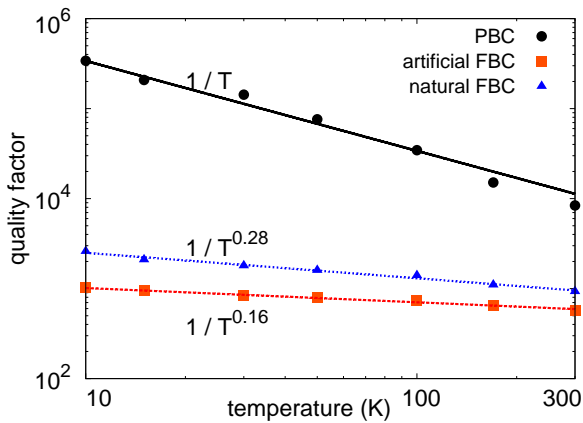}}
  \end{center}
  \caption{(Color online) Quality factor versus temperature for Sample-A with PBC (black filled circle), FBC and artificial actuation (red filled square), and FBC and natural actuation (blue filled trigon).}
  \label{fig_qfactor}
\end{figure}
\begin{figure}[htpb]
  \begin{center}
    \scalebox{0.75}[0.75]{\includegraphics[width=8cm]{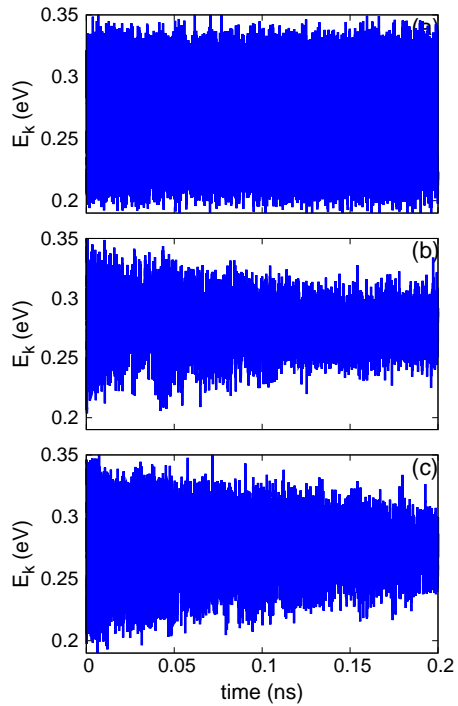}}
  \end{center}
  \caption{(Color online) Kinetic energy time history at 10 K for zigzag graphene nanoresonator of size $(L_{x}, L_{y}) = (19.7, 29.82)$~{\AA} with (a) PBC, (b) FBC and artificial actuation, and (c) FBC and natural actuation.}
  \label{fig_ek_ga87}
\end{figure}
The flipping phenomenon is a disaster for the quality factor of the mechanical resonator, because the positions of atoms in the edge region are suddenly relocated during each flipping. This is extremely harmful for the coherence of the mechanical oscillation, leading to very low quality factors. Panels (a) and (b) in Fig.~\ref{fig_ek_300K_ga830} are simulation results for Sample-A at 300 K, which are similar to that at 10 K. The energy damping for the nanoresonator with FBC is obviously much stronger than that with PBC. Fig.~\ref{fig_qfactor} shows the temperature dependence for the quality factor. It shows that the quality factor for the nanoresonator with FBC (red filled squares) is about three orders smaller than the nanoresonator with PBC (black filled circles) in whole temperature range. The quality factor for the nanoresonator with PBC is inversely proportional to the temperature as $1/T$. This temperature behavior results from the three-phonon scattering,\cite{Holland} which is the only intrinsic decaying mechanism in the graphene nanoresonator with PBC. For nanoresonators with FBC, the power factor in the temperature dependence is considerably reduced to be only 0.16. It is because the imaginary edge vibration effect (flipping process) is not sensitive to the temperature. The combination of the imaginary edge vibration effect and phonon-phonon scattering effect results in a power factor smaller than 1. A small power factor of 0.16 indicates that the imaginary edge vibration effect overcomes the phonon-phonon scattering effect.

\begin{figure}[htpb]
  \begin{center}
    \scalebox{0.75}[0.75]{\includegraphics[width=8cm]{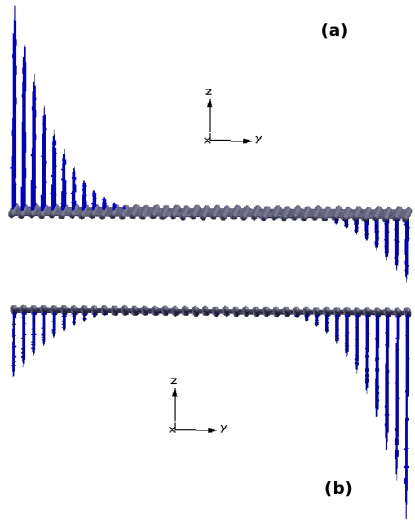}}
  \end{center}
  \caption{(Color online) Two edge vibrations localized nearby the two free short edges in armchair graphene nanoresonator of size $(L_{x}, L_{y}) = (25.56, 49.2)$~{\AA}. Vibrational amplitudes decay exponentially from edge into center. The frequencies of these two edge vibrations in (a) and (b) are degenerate at 3.02~{cm$^{-1}$}.}
  \label{fig_u_edgemode_gz620}
\end{figure}
\begin{figure}[htpb]
  \begin{center}
    \scalebox{1.0}[1.0]{\includegraphics[width=8cm]{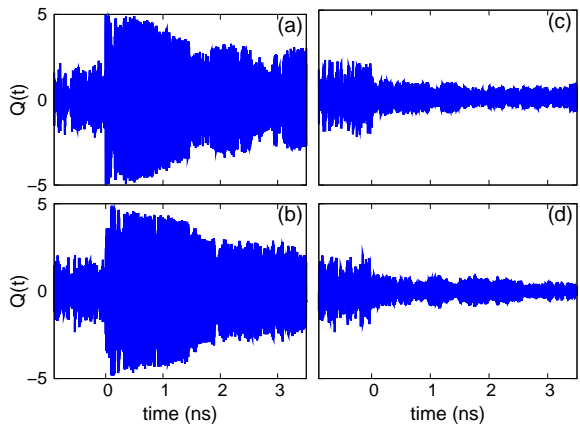}}
  \end{center}
  \caption{(Color online)  Normal mode coordinate time history of the two edge vibrations in Fig.~\ref{fig_u_edgemode_gz620} for armchair graphene nanoresonator of size $(L_{x}, L_{y}) = (25.56, 49.2)$~{\AA} with FBC at 10~{K}. (a) and (b) are for these two edge vibrations in artificial resonator. (c) and (d) are for natural resonator.}
  \label{fig_normalmode_gz620}
\end{figure}
\begin{figure}[htpb]
  \begin{center}
    \scalebox{0.75}[0.75]{\includegraphics[width=8cm]{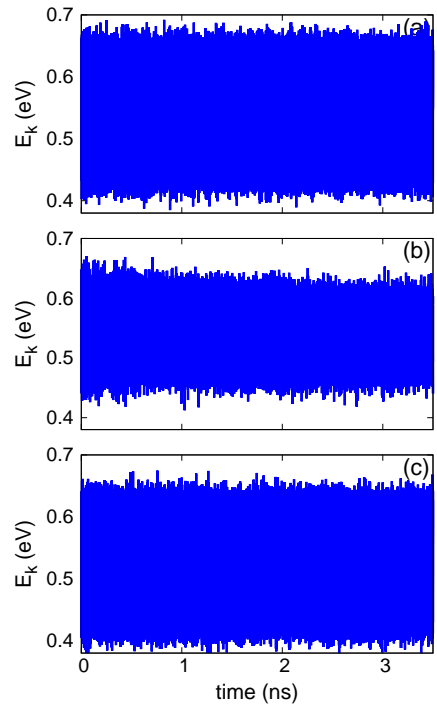}}
  \end{center}
  \caption{(Color online) Kinetic energy time history at 10 K for armchair graphene resonator of $(L_{x}, L_{y}) = (25.56, 49.2)$~{\AA} with (a) PBC, (b) FBC and artificial actuation, and (c) FBC and natural actuation.}
  \label{fig_ek_gz620}
\end{figure}
Fig.~\ref{fig_ek_ga87} shows time history of kinetic energy for zigzag graphene nanoresonator of another size $(L_{x}, L_{y}) = (19.7, 29.82)$~{\AA} at 10~{K}. There are also two imaginary edge vibrations at the free zigzag edges of this graphene nanoribbon. The frequencies of these imaginary edge vibrations are almost the same as that of Sample-A, because edge vibrations are only sensitive to the configuration of edge regions which are the same in both systems. We observe similar phenomenon in this nanoresonator. Panel (a) shows that there is only weak energy damping in the nanoresonator with PBC, resulting in a very high quality factor. Panel (b) shows much stronger energy damping for the nanoresonator with FBC, which leads to an extremely low quality factor.

In the above, we find imaginary edge vibrations in zigzag graphene nanoresonators with FBC. Our simulations have disclosed a fact that the flipping processes of these imaginary edge vibrations break the coherence of the resonator's mechanical oscillation, which leads to extremely low quality factors in zigzag graphene nanoresonators with FBC. It is natural to ask what will be the quality factor for graphene nanoresonators without imaginary edge vibration. If our above analysis based on the flipping processes of imaginary edge vibrations are true, we should obtain weak energy damping and high quality factors for resonators without imaginary edge vibration. To confirm this assumption, we carry out simulation for graphene nanoresonators with armchair edges, where there is no imaginary edge vibration. The first armchair graphene nanoresonator we studied is of size $(L_{x}, L_{y}) = (25.56, 49.2)$~{\AA}. Fig.~\ref{fig_u_edgemode_gz620} display the two lowest-energy edge vibrations for FBC. The vibrational amplitude also shows an exponential decaying from edge into center region, which is a characteristic of localized vibrations. These two edge vibrations are degenerate at 3.02~{cm$^{-1}$} because of two free edges, and their eigen vectors happen to mix together. Normal mode coordinates of these two edge vibrations at 10~{K} are demonstrated in Fig.~\ref{fig_normalmode_gz620}~(a) and (b).
\begin{figure}[htpb]
  \begin{center}
    \scalebox{0.7}[0.7]{\includegraphics[width=8cm]{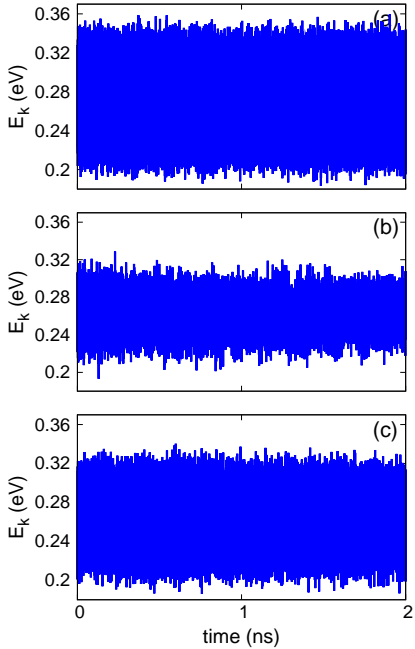}}
  \end{center}
  \caption{(Color online) Kinetic energy time history at 10 K for armchair graphene nanoresonator of $(L_{x}, L_{y}) = (25.56, 24.6)$~{\AA} with (a) PBC, (b) FBC and artificial actuation, and (c) FBC and natural actuation.}
  \label{fig_ek_gz610}
\end{figure}
 Obviously, there is no flipping phenomenon anymore. Instead, these two edge vibrations involve in continuous vibrations, which helps to preserve the coherence of the resonator's mechanical oscillation. As a result, we should observe very weak energy damping. Indeed, Fig.~\ref{fig_ek_gz620} shows that the energy damping for the mechanical oscillation is very weak in resonators with both PBC (in panel (a)) and FBC (in panel (b)). It means that the quality factor is not necessarily low for graphene nanoresonators even if FBC is applied. These results provide an indirect evidence for the correctness of our analysis of attributing low quality factor to the flipping process of imaginary edge vibrations. Fig.~\ref{fig_ek_gz610} shows similar kinetic energy time history for nanoresonator with armchair edges of another size ($(L_{x}, L_{y}) = (25.56, 24.6)$~{\AA}) at 10~{K}.

\subsection{effects from artificial or natural actuation}
Up to now, we have revealed one underlying mechanism for extremely low quality factors in zigzag graphene nanoresonators with free edges, where the flipping processes of imaginary edge vibrations play an important role. We have also shown that high quality factors can be obtained for armchair
\begin{figure}[htpb]
  \begin{center}
    \scalebox{1.0}[1.0]{\includegraphics[width=8cm]{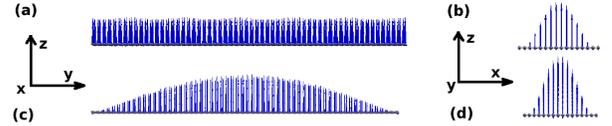}}
  \end{center}
  \caption{(Color online) (a) and (b) are two different side views for the first natural vibration (non-edge vibration) of Sample-A with PBC. It is only sinuous function in $x$ direction. (c) and (d) are two different side views for the first natural vibration (non-edge vibration) of Sample-A with FBC. It is sinuous function in both $x$ and $y$ directions.}
  \label{fig_u_mode1_ga830}
\end{figure}
\begin{figure}[htpb]
  \begin{center}
    \scalebox{1.0}[1.0]{\includegraphics[width=8cm]{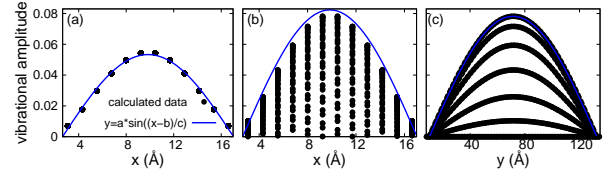}}
  \end{center}
  \caption{The fitting of the first natural vibrations to sinusoidal function $y=a*\sin (x-b)/c$ for Sample-A. (a). is for the distribution of the vibration in x direction with periodic boundary condition applied in y direction, resulting in $a=0.053$, $b=2.64$~{\AA}, and $c=4.58$~{\AA}. (b). is for the distribution of the vibration in x direction with free boundary condition applied in y direction, resulting in $a=0.082$, $b=2.64$~{\AA}, and $c=4.58$~{\AA}. (c). is for the distribution of the vibration in y direction with free boundary condition applied in y direction, resulting in $a=0.078$, $b=10.35$~{\AA}, and $c=38.91$~{\AA}.}
  \label{fig_u_fit_sin}
\end{figure}
\begin{figure}[htpb]
  \begin{center}
    \scalebox{0.75}[0.75]{\includegraphics[width=8cm]{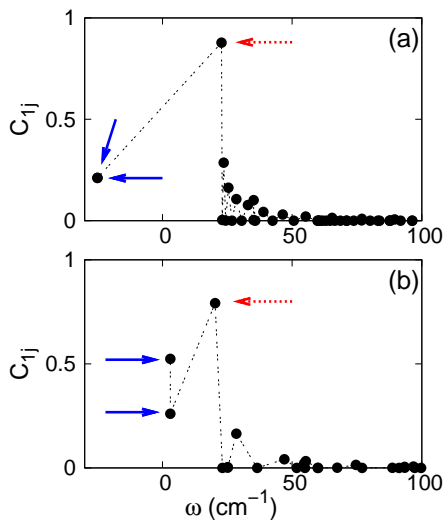}}
  \end{center}
  \caption{(Color online) (a) Decomposition of the first natural vibration of Sample-A with PBC ($\xi_{1}$) in terms of the natural vibrations of Sample-A with FBC ($\chi_{j}$, j=1,2,3,...,$3N$): $C_{1j}=\langle \xi_{1} | \chi_{j} \rangle$. Two blue solid arrows indicate the two edge vibrations shown in Fig.~\ref{fig_u_edgemodemode_ga830} for Sample-A with FBC. The imaginary value of the frequency $\omega=\pm 25i$cm$^{-1}$ has been intensively shown as a negative value $\omega=-25$cm$^{-1}$. The red dotted arrow indicates the first natural vibration of Sample-A with FBC ($\chi_{1}$). (b) Similar decomposition for the first natural vibration of armchair graphene nanoresonator of $(L_{x}, L_{y}) = (25.56, 49.2)$~{\AA} with PBC. Two blue solid arrows indicate the two edge vibrations shown in Fig.~\ref{fig_u_edgemode_gz620}.}
  \label{fig_decompose}
\end{figure}
 graphene nanoresonators with free edges, because of the absence of imaginary edge vibrations. Now we switch to the discussion of another effect from free edges in graphene nanoresonators. From a basic point of view, the mechanical oscillation of a natural resonator is actuated by mechanically exciting a particular natural vibration of the system. After the actuation of the resonator, the frequency of the resonator equals to the frequency of the natural vibration, and the mechanical oscillation of the resonator corresponds to the vibrational morphology of the natural vibration. Let's review how we actuated nanoresonators in the above. In all previous simulations, nanoresonators are actuated by adding a velocity $\Delta \vec{v}_{j}\propto\vec{e}_{z} \sin(x_{j}/L_{x})$ to each atom $j$. Fig.~\ref{fig_u_mode1_ga830}~(a) and (b) show that this sinuous function is a natural vibration in Sample-A with PBC. We use $\xi_{j}$ (with $j=1,2,3,...,3N$) to denote eigen vectors of all natural vibrations in graphene nanoresonators with PBC; particularly, we use $\xi_{1}=\sin(x/L_{x})$. We use a sinusoidal function $y=a\sin (x-b)/c$ to fit $\xi_{1}$ mode in Fig.~\ref{fig_u_fit_sin}~(a), where the relative fitting errors for parameters ($a$, $b$ and $c$) are less than $0.1\%$. This vibration does not depend on the y coordinate, so all data falls in one sinusoidal function. It should be noted that the sinuous function is the natural vibration in a rectangular plate, while the Bessel function is the natural vibration in the circular plate (such as the circular graphene in Ref.~\onlinecite{KimSYapl}).\cite{Polyanin} The graphene nanoresonator with PBC is a natural resonator if its mechanical oscillation is actuated according to $\xi_{1}$. However, the sinuous function is not a natural vibration in Sample-A with FBC. We use $\chi_{j}$ (with $j=1,2,3,...,3N$) to denote eigen vectors of all natural vibrations in Sample-A with FBC. The lowest-energy natural vibration (non-edge vibration) is shown in Fig.~\ref{fig_u_mode1_ga830}~(c) and (d) for Sample-A with FBC.
\begin{figure}[htpb]
  \begin{center}
    \scalebox{1.0}[1.0]{\includegraphics[width=8cm]{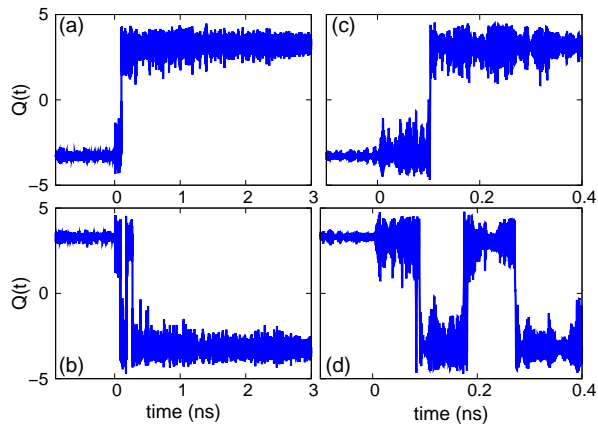}}
  \end{center}
  \caption{(Color online) Normal mode coordinate time history of the two edge vibrations in Fig.~\ref{fig_u_edgemodemode_ga830} showing flipping phenomena. The simulated system is Sample-A at 10~{K} with FBC and natural actuation. (a) is for the edge vibration shown in Fig.~\ref{fig_u_edgemodemode_ga830}~(a). (b) is for the edge vibration shown in Fig.~\ref{fig_u_edgemodemode_ga830}~(b). (c) Close-up for the flipping phenomenon in the first edge vibration. (d) Close-up for the flipping phenomenon in the second edge vibration.}
  \label{fig_normalmode_natural_ga830}
\end{figure}
 We denote this particular natural vibration as $\chi_{1}$. Fig.~\ref{fig_u_fit_sin}~(b) and (c) are the fitting for the distribution of vibration of $\chi_{1}$. The fitting curve in panel (b) is for the those atoms in the middle of the graphene nanoribbon with $y\approx70$~{\AA}, while the fitting curve in panel (c) is for the those atoms in the middle of the graphene nanoribbon with $x\approx10$~{\AA}. It clearly shows that this natural vibration is a sinuous function in both $x$ and $y$ directions. Hence, the graphene nanoresonator with FBC is not a natural resonator if its mechanical oscillation is actuated according to $\xi_{1}$. We name this type of resonator as an artificial resonator, which is actuated according to an unnatural vibration of the system. For an artificial resonator, many natural vibrations are excited simultaneously during the actuation of the mechanical oscillation of the nanoresonator. $\xi_{1}$ can be decomposed in terms of $\chi_{j}$ (with $j=1,2,3,...,3N$):
\begin{equation}
\xi_{1}=\sum_{j=1}^{3N}c_{1j}\chi_{j}.
\label{eq_decompose}
\end{equation}
Those vibrations with large coefficients $|c_{1j}|$ are considerably excited during the actuation of the nanoresonator according to $\xi_{1}$. Fig.~\ref{fig_decompose}~(a) displays the coefficient $c_{1j}$ for all natural vibrations in Sample-A with FBC. Those natural vibrations above 100 cm$^{-1}$ are not excited. The highest peak (indicated by red doted arrow) corresponds to the natural vibration $\chi_{1}$ in Sample-A with FBC. We find that the other two strong peaks (indicated by blue solid arrows) are with respect to the two imaginary edge vibrations at the free edges as shown Fig.~\ref{fig_u_edgemodemode_ga830}. Actually, from the vibrational morphology of natural vibrations $\xi_{1}$, $\chi_{1}$, and the two imaginary edge vibrations $\chi_{\rm edge vibration}$, the relationship among them can be simply obtained: $\xi_{1}\approx\chi_{1}+\chi_{\rm edge vibration}$. This decomposition indicates that the two edge vibrations are also considerably excited when the nanoresonator with FBC is actuated according to $\xi_{1}$. These edge vibrations are not in thermal equilibrium state. As a result, the edge effect is stronger in artificial resonator than the natural resonator.

The imaginary edge vibration effect (discussed in previous section) and the artificial effect are both important mechanisms for low quality factors in graphene nanoresonators with FBC. We can split these two effects separately. There is no artificial effect if the graphene nanoresonator with FBC is actuated naturally according to $\chi_{1}$. This is a natural mechanical resonator, so it can be used to study the pure imaginary edge vibration effect. Fig.~\ref{fig_ek_fbc_ga830}~(b) shows the time history of kinetic energy in Sample-A with FBC, which is naturally actuated according to its $\chi_{1}$. The energy damping is still very strong, which is only slightly smaller than the artificial resonator shown in panel (a) of the same figure. The normal mode coordinates for the two imaginary edge vibrations in Sample-A are monitored in Fig.~\ref{fig_normalmode_natural_ga830}. We observe the same flipping phenomenon as that in the artificial resonator shown in Fig.~\ref{fig_normalmode_artificial_ga830}. As we have discussed above, the flipping process breaks the coherence of the mechanical oscillation, leading to a low quality factor. The strong energy damping in Fig.~\ref{fig_ek_fbc_ga830}~(b) means that the imaginary edge vibration effect is the key mechanism for low quality factors; while the small difference between Fig.~\ref{fig_ek_fbc_ga830}~(b) and (a) indicates that the artificial effect is only of minor importance for the reduction of quality factor. The quality factor is mainly determined by the imaginary edge vibration effect in graphene nanoresonators with imaginary edge vibrations. Fig.~\ref{fig_qfactor} shows that the quality factor in the natural resonator is slightly larger than the artificial resonator. The temperature dependence power factor in the natural resonator is 0.28, which is larger than the value of 0.16 for the artificial resonator. This slight difference is due to the artificial effect. The kinetic energy time history for zigzag graphene natural resonator with FBC of another size is shown in Fig.~\ref{fig_ek_ga87}~(c), which confirms that the imaginary edge vibration effect is the major reason for low quality factors.

We have previously shown that there is no imaginary edge vibration in the armchair graphene nanoresonator with FBC, so it can be used to study the pure artificial effect on the quality factor. Fig.~\ref{fig_ek_gz620}~(c) shows the kinetic energy time history for armchair graphene nanoresonator with FBC of size $(L_{x}, L_{y}) = (25.56, 49.2)$~{\AA}. The mechanical oscillation of the nanoresonator is actuated naturally according to $\chi_{1}$. The result displays very high quality factor in this natural mechanical resonator, although the FBC has been applied. The minimum energy damping is as weak as the mechanical resonator with PBC shown in panel (a). Fig.~\ref{fig_normalmode_gz620}~(c) and (d) monitor the time history of normal mode coordinates of the two edge vibrations in this system. First of all, there is no flipping phenomenon due to the absence of imaginary edge vibrations. Furthermore, $Q(t)$ is further suppressed after the heat bath is removed at $t=0$, which indicates that these two edge vibrations are quite difficult to be excited if one does not excite them artificially. Fig.~\ref{fig_normalmode_gz620}~(a) and (b) show that they will be considerably excited if the mechanical oscillation of the graphene nanoresonator is acuated artificially according to $\xi_{1}$ which can be decomposed in terms of $\chi_{k}$ in Fig.~\ref{fig_decompose}~(b). Fig.~\ref{fig_ek_gz610}~(c) shows same results in armchair graphene nanoresonator with FBC of another size $(L_{x}, L_{y}) = (25.56, 24.6)$~{\AA}, which is also actuated naturally.

Based on the two edge effects discussed above, we can give some explanations for the experiment. In the experiment,\cite{Zande} it was found that edge effects are important for quality factors of graphene nanoresonators. For samples with free edges, the measured quality factors are low and the values vary between identically prepared samples. We can learn from our simulations that the imaginary edge vibrations are of key importance for low quality factors. Another experiment has shown that it is easily for free zigzag edges in the graphene nanoribbon to undergo a reconstruction and become more stable.\cite{GassMH,JiaX} The reconstruction is attributed to the edge vibrations at the free edge;\cite{ShenoyVB,Engelund} thus it will partially eliminate the imaginary edge vibrations. Considering the later experiment, a reasonable explanation for the experiment in Ref.~\onlinecite{Zande} is that some samples happen to reconstruct their free edges, which help to remove some imaginary edge vibrations. As a result, the measured quality factor will be higher in these samples undergoing reconstruction, while quality factors are still low in samples without reconstruction. This explains the variation of quality factors between identically prepared graphene nanoresonators.

\section{conclusion}
Our molecular dynamics simulations have disclosed two effects from the free edges on the graphene nanoresonator, which are both responsible for low quality factors. They are the imaginary edge vibration effect and the artificial effect. Imaginary edge vibrations exist at free zigzag edges, so the imaginary edge vibration effect is of crucial importance for low quality factors in zigzag graphene nanoresonators; while the artificial effect only slightly further reduces quality factors. The flipping process for imaginary edge vibrations between two doubly degenerate warping states will seriously break the coherence of the mechanical oscillation of the graphene nanoresonator. The power factor in the temperature dependence of quality factors is 1.0 for graphene nanoresonator with PBC, which reflects the intrinsic phonon-phonon scattering mechanism for the energy damping. This power factor decreases to be 0.28 in natural nanoresonators with FBC. This reduction is due to the flipping phenomenon for the imaginary edge vibrations. It is further reduced to be 0.16 in artificial nanoresonators with FBC. This further reduction results from the artificial effect.

There is no imaginary edge vibration in armchair graphene nanoresonator with FBC, so the artificial effect is the only mechanism for the decrease of quality factors. An extremely high quality factor can be obtained for armchair graphene nanoresonator with FBC, if its mechanical oscillation is naturally actuated. This result indicates that free edges not necessarily lead to low quality factors for graphene nanoresonators. Our simulations provide advices for experimentalists that high quality factors can be obtained if graphene nanoresonators with FBC are of armchair edges and actuated naturally.

\textbf{Acknowledgements} We thank Prof. B. S. Wang at IOS-CAS for insightful discussions, and S. Liu for useful discussions, and H. N. Li and H. L. Zhou for calculation resource. The work is supported by a URC grant of R-144-000-257-112 of National University of Singapore.


\begin{thebibliography}{}
\bibitem{BunchJS} J. S. Bunch, A. M. van der Zande, S. S. Verbridge, I. W. Frank, D. M. Tanenbaum, J. M. Parpia, H. G. Craighead, and P. L. McEuen, Science \textbf{315}, 490 (2007).

\bibitem{Zande} A. M. van der Zande, R. A. Barton, J. S. Alden, C. S. Ruiz-Vargas, W. S. Whitney, P. H. Q. Pham, J. Park, J. M. Parpia, H. G. Craighead, and P. L. McEuen, Nano. Lett. \textbf{10}, 4869 (2010).

\bibitem{SanchezDG} D. G. Sanchez, A. M. van der Zande, A. San Paulo, B. Lassagne, P. L. McEuen, and A. Bachtold, Nano. Lett. \textbf{8}, 1399 (2008).

\bibitem{SeoanezC} C. Seoanez, F. Guinea, and A. H. Castro Neto, Phys. Rev. B \textbf{76}, 125427 (2007).

\bibitem{AtalayaJ} J. Atalaya, A. Isacsson, and J. M. Kinaret, Nano. Lett. \textbf{8}, 4196 (2008).

\bibitem{JiangJW} J.-W. Jiang and J.-S. Wang, arXiv:1108.5805v1 (2011).

\bibitem{KimSY} S. Y. Kim and H. S. Park, Nano. Lett. \textbf{9}, 969 (2009).

\bibitem{GassMH} M. H. Gass, U. Bangert, A. L. Bleloch, P. Wang, R. R. Nair, and A. K. Geim, Nat. Nanotechnol. \textbf{3}, 676 (2008).

\bibitem{ShenoyVB} V. B. Shenoy, C. D. Reddy, A. Ramasubramaniam, and Y. W. Zhang, Phys. Rev. Lett. \textbf{101}, 245501 (2008).

\bibitem{JiaX} X. Jia, M. Hofmann, V. Meunier, B. G. Sumpter, J. Campos-Delgado, J. M. Romo-Herrera, H. Son, Y. P. Hsieh, A. Reina, J. Kong, M. Terrones, and M. S. Dresselhaus, Science, \textbf{323}, 1701 (2009).

\bibitem{Engelund} M. Engelund, J. A. Furst, A. P. Jauho, and M. Brandbyge, Phys. Rev. Lett. \textbf{104}, 036807 (2010).

\bibitem{Brenner} D. W. Brenner, O. A. Shenderova, J. A. Harrison, S. J. Stuart, B. Ni and S. B. Sinnott, J. Phys.:Condens. Matter \textbf{14}, 783 (2002).

\bibitem{Nose} S. No\'se, J. Chem. Phys. \textbf{81}, 511 (1984).

\bibitem{Hoover} W. G. Hoover, Phys. Rev. A, \textbf{31}, 1695 (1985).

\bibitem{LanJ} J. Lan, J.-S. Wang, C. K. Gan, and S. K. Chin, Phys. Rev. B \textbf{79}, 115401 (2009).

\bibitem{DaviesPCW} P. C. W. Davies, \textit{The forces of nature}, (Cambridge University Press, New York, 1986).

\bibitem{NagaosaN} N. Nagaosa, \textit{Quantum field theory in condensed matter physics}, (Spring, Berlin, 1999).

\bibitem{MorandiG} G. Morandi, \textit{Statistical mechanics : an intermediate course}, (World Scientific, Singapore, 1995).

\bibitem{GoldstoneJ} J. Goldstone, A. Salam, and S. Weinberg, Phys. Rev. \textbf{127}, 965 (1962).

\bibitem{Holland} M. G. Holland, Phys. Rev. \textbf{134}, A471 (1964).

\bibitem{KimSYapl} S. Y. Kim and H. S. Park, Appl. Phys. Lett. \textbf{94} 101918 (2009).

\bibitem{Polyanin} A. D. Polyanin, \textit{Handbook of Linear Partial Differential Equations for Engineers and Scientists} (CRC Press/C$\&$H, Florida,) (2002).
\end{thebibliography}
\end{document}